\documentclass[12pt]{article}
\usepackage{color}
\usepackage{a4}
\usepackage{latexsym}
\usepackage{cite}
\usepackage{bm}
\usepackage{lineno}

\usepackage{array}
\newcolumntype{T}{>{\ttfamily} c}
\newcolumntype{M}{>{$\displaystyle} c <{$}}

\usepackage{epsfig}
\usepackage{graphicx}

\usepackage{pslatex}
\usepackage[latin1]{inputenc}
\usepackage[T1]{fontenc}

\newcommand{\gsim}{\raisebox{-0.7mm}{$\:\:\stackrel{>}{{\scriptstyle
 \sim}}\:\: $} }

\newcommand{\Qeq}{\raisebox{-0.1mm}{$\:\!\stackrel{Q_{\phantom{i\!\!\!\!}}}{=}\:\:$}}

\newcommand{\beq}{\begin{equation}}
\newcommand{\eeq}{\end{equation}}
\newcommand{\bea}{\begin{eqnarray}}
\newcommand{\eea}{\end{eqnarray}}
\newcommand{\nn}{\nonumber}

\newcommand{\ra}{\rightarrow}

\newcommand{\als}{\alpha_{\rm s}}
\newcommand{\ars}{a_{\rm s}}
\newcommand{\ep}{\varepsilon}

\newcommand{\hspp}{{\hspace{5mm}}}

\def\dNnum#1{\delta_{\:\!N,#1}}

\def\frct#1#2{\mbox{\small{$\displaystyle\frac{#1}{#2}$}}}

\def\as(#1){{\alpha_{\rm s}^{\,#1}}}
\def\ar(#1){{a_{\rm s}^{\,#1}}}

\def\zr#1{{\zeta_{#1}^{}}}
\def\mus{{\mu^{\,2}}}

\def\B(#1,#2){{\beta_{#1}^{\,#2}}}

\def\ncs{{n_{c}^{\,2}}}

\def\ca{{C^{}_A}}
\def\cas{{C^{\,2}_A}}
\def\cat{{C^{\,3}_A}}
\def\caf{{C^{\,4}_A}}
\def\cf{{C^{}_F}}
\def\cfs{{C^{\, 2}_F}}
\def\cft{{C^{\, 3}_F}}
\def\cff{{C^{\, 4}_F}}
\def\nf{{n^{}_{\! f}}}
\def\nfz{{n^{\,0}_{\! f}}}
\def\nfo{{n^{\,1}_{\! f}}}
\def\nfs{{n^{\,2}_{\! f}}}
\def\nft{{n^{\,3}_{\! f}}}

\def\DNn#1{D_0^{\:#1}}
\def\DNm#1{D_{-1}^{\:#1}}
\def\DNp#1{D_1^{\:#1}}
\def\DNpp#1{D_2^{\:#1}}

\def\etaD#1{\eta^{\,#1}}
\def\nuD#1{\nu^{\,#1}}

\def\xm1{{(1 \! - \! x)}}
\def\xp1{{(1 \! + \! x)}}

\def\Lnt(#1){\ln^{\,#1}(1\!-\!x)}
\def\pqq(#1){p_{\rm{qq}}(#1)}

\def\S(#1){{{S}_{#1}}}
\def\Ss(#1,#2){{{S}_{#1,#2}}}
\def\Sss(#1,#2,#3){{{S}_{#1,#2,#3}}}
\def\Ssss(#1,#2,#3,#4){{{S}_{#1,#2,#3,#4}}}
\def\Sssss(#1,#2,#3,#4,#5){{{S}_{#1,#2,#3,#4,#5}}}
\def\Ssssss(#1,#2,#3,#4,#5,#6){{{S}_{#1,#2,#3,#4,#5,#6}}}
\def\Sssssss(#1,#2,#3,#4,#5,#6,#7){{{S}_{#1,#2,#3,#4,#5,#6,#7}}}

\def\Sp(#1,#2){{{S}_{#1}^{\,#2}}}

\def\H(#1){{\rm{H}}_{#1}}
\def\Hh(#1,#2){{\rm{H}}_{#1,#2}}
\def\Hhh(#1,#2,#3){{\rm{H}}_{#1,#2,#3}}
\def\Hhhh(#1,#2,#3,#4){{\rm{H}}_{#1,#2,#3,#4}}
\def\Hhhhh(#1,#2,#3,#4,#5){{\rm{H}}_{#1,#2,#3,#4,#5}}
\def\Hhhhhh(#1,#2,#3,#4,#5,#6){{\rm{H}}_{#1,#2,#3,#4,#5,#6}}

\begin{document}
\setlength{\parskip}{0.2cm}
\setlength{\baselineskip}{0.55cm}

\begin{titlepage}
\noindent
DESY 18--072 \hfill May 2018\\
Nikhef 2018-023 \\
LTH 1165 \\
\vspace{0.6cm}
\begin{center}
{\LARGE \bf On quartic colour factors in splitting functions\\[1ex]
 and the gluon cusp anomalous dimension}\\ 
\vspace{1.4cm}
\large
S. Moch$^{\, a}$, B. Ruijl$^{\, b}$, T. Ueda$^{\, c}$, 
J.A.M. Vermaseren$^{\, d}$ and A. Vogt$^{\, e}$\\
\vspace{1.2cm}
\normalsize
{\it $^a$II.~Institute for Theoretical Physics, Hamburg University\\
\vspace{0.5mm}
Luruper Chaussee 149, D-22761 Hamburg, Germany}\\
\vspace{4mm}
{\it $^b$Institute for Theoretical Physics, ETH Z\"urich\\
\vspace{0.5mm}
Wolfgang-Pauli-Str.~27, 8093 Z\"urich, Switzerland}\\
\vspace{4mm}
{\it $^c$ Department of Materials and Life Science, Seikei University\\
\vspace{0.5mm}
3-3-1 Kichijoji Kitamachi, Musashino-shi, Tokyo 180-8633, Japan}\\
\vspace{4mm}
{\it $^d$Nikhef Theory Group \\
\vspace{0.5mm}
Science Park 105, 1098 XG Amsterdam, The Netherlands} \\
\vspace{4mm}
{\it $^e$Department of Mathematical Sciences, University of Liverpool\\
\vspace{0.5mm}
Liverpool L69 3BX, United Kingdom}\\
\vspace{1.4cm}
{\large \bf Abstract}
\vspace{-0.2cm}
\end{center}
We have computed the contributions of the quartic Casimir invariants to the
four-loop anomalous dimensions of twist-2 spin-$N$ operators at $N \leq 16$.
The results provide new information on the structure of the 
next-to-next-to-next-to-leading order (N$^3$LO) splitting functions 
$P_{\rm ik}^{\,(3)}(x)$ for the evolution of parton distributions, and 
facilitate approximate expressions which include the quartic-Casimir 
contributions to the (light-like) gluon cusp anomalous dimension.
These quantities turn out to be closely related, by a generalization of the
lower-order `Casimir scaling', to the corresponding quark results.
Using these findings, we present an approximate result for the four-loop gluon 
cusp anomalous dimension in QCD which is sufficient for phenomenological 
applications.

\vspace*{0.2cm}
\end{titlepage}

%
\section{Introduction}
\label{sec:intro}
\vspace*{-1mm}

Over the past years,
the next-to-next-to-leading order (NNLO, N$^2$LO) of perturbative QCD has 
become the standard approximation for many hard-scattering processes at the
LHC and other high-energy colliders.
In certain cases, e.g., when a very high accuracy is required or when the
NNLO corrections are rather large, it is useful to extend the analyses to 
the next order,~N$^3$LO.
Coefficient functions (partonic cross sections) have been computed at N$^3$LO
for inclusive lepton-hadron deep-inelastic scattering (DIS) \cite{MVV610+} 
and Higgs production in proton-proton collisions \cite{Higgs1,Higgs2}. 
Very recently, first N$^3$LO results have been presented for jet production 
in DIS \cite{N3LOjet}.

In principle, N$^3$LO analyses of processes involving initial-state hadrons 
require parton distribution functions (PDFs) evolved with the four-loop 
splitting functions.
These functions also include quantities that are relevant beyond the evolution 
of PDFs. 
In particular, their leading behaviour for large momentum fractions $x$ is 
given by important universal quantities, the (light-like) cusp anomalous 
dimensions of quarks and gluons \cite{Korch89}.
The complete computation of the four-loop splitting functions is a formidable
task. Until now, a phenomenologically relevant amount of partial results has 
been published only for the (non-singlet) quark-quark splitting functions 
\cite{DRUVV,MRUVV1}.

In this letter, we address a specific part of the four-loop 
flavour-singlet splitting functions, the terms with quartic Casimir invariants 
which occur at this order for the first time.
As~shown below, the present partial results for these terms provide
structural and numerical information that is relevant for future research
on N$^3$LO corrections and for QCD phenomenology beyond PDFs.

%
\section{Notations and general properties}
\label{sec:genprops}
\setcounter{equation}{0}
\vspace*{-1mm}

The QCD evolution equations for the flavour-singlet quark and gluon 
distributions of hadrons,
\beq
\label{sgPDFs}
  q_{\rm s}^{}(x,\mus) \; = \; \sum_{i=1}^{n_{\!f}} \left[\,
  q_i^{}(x,\mus) + \,\bar{q}_i^{}(x,\mus) \:\!\right] 
\quad \mbox{and} \quad g(x,\mus)
\; , 
\eeq
can be written as
\beq
\label{sgEvol}
  \frac{d}{d \ln\mus} \; 
  \Big( \begin{array}{c} \! q_{\rm s}^{} \!\! \\ \!g\!  \end{array} \Big)
  \: = \: \left( 
  \begin{array}{cc} \! P_{\rm qq} & P_{\rm qg} \!\!\! \\
                    \! P_{\rm gq} & P_{\rm gg} \!\!\! \end{array} \right) 
  \otimes
  \Big( \begin{array}{c} \!q_{\rm s}^{}\!\! \\ \!g\!  \end{array} \Big)
  \:\: .
\eeq
Here $q_i^{}(x,\mus)$, $\bar{q}_i^{}(x,\mus)$ and $g(x,\mus)$ denote
the respective number distributions of quarks and antiquarks of flavour $i$ 
and of the gluons in the fractional hadron momentum $x$,
$\otimes$ stands for the Mellin convolution in the momentum variable,
and $\mu$ represents the factorization scale. 
In the present context, the renormalization scale can be identified with 
$\mu$ without loss of information.

The quark-quark splitting function $P_{\rm qq}$ can be expressed as 
$ P_{\rm qq} = P_{\rm ns}^{\,+} + P_{\rm ps}^{} $ 
in terms of the non-singlet splitting function $P_{\rm ns}^{\,+}$ for 
quark-antiquark sums and a pure-singlet contribution $P_{\rm ps}^{}$. 
The~splitting functions can be expanded in powers of the strong coupling 
constant,
\beq
\label{Pexp}
 P_{\rm ik}^{}\left(x,\als\right) \; = \; \sum_{n=0}\,
  \ar(n+1) P^{\,(n)}_{\rm ik}(x)
\quad \mbox{with} \quad \ars \:\equiv\; \frct{\als(\mus)}{4\pi}
  \:\: .
\eeq
The off-diagonal quantities $P_{\rm qg}$ and $P_{\rm gq}$ include integrable
logarithms up to $\ar(n)\,\ln^{\,2n-2\!}\xm1$ in the threshold limit $x \ra 1$, 
see refs.~\cite{OffDxto1}, while the diagonal quantities $P_{\rm qq}$ and 
$P_{\rm gg}$ have the form \cite{DMS05}
\beq
\label{xto1}
  P_{\rm kk}^{\,(n-1)}(x) \;=\;
        \frac{x\,A_{n,\rm k}}{(1-x)_+}
  \,+\, B_{n,\rm k} \, \delta \xm1
  \,+\, C_{n,\rm k} \, \ln \xm1
  \,+\, D_{n,\rm k} \,+\, {\cal O}\big( \xm1 \ln^{\:\!\ell\!} \xm1 \big)
\; ,
\eeq
where $A_{n,\rm q}$ and $A_{n,\rm g}$ are the (light-like) $n$-loop quark and 
gluon cusp anomalous dimensions \cite{Korch89}. The~coefficients $C_{n,\rm k}$
and $D_{n,\rm k}$ can be predicted from lower-order information 
\cite{DMS05,MRUVV1}. 
In the small-$x$ (high-energy, BFKL \cite{BFKL1}) limit, the splitting 
functions are single-logarithmic enhanced with terms up to 
$x^{\,-1}\ln^{\,n} x$ for $P_{\rm gk}^{\,(n)}$ and 
$x^{\,-1}\ln^{\,n-1} x$ for $P_{\rm qk}^{\,(n)}$ \cite{BFKL2}.
 
The splitting functions in eq.~(\ref{sgEvol}) are related to the anomalous 
dimensions of twist-2 spin-$N$ operators with $N = 2,\, 4,\, 6,\, \ldots$ 
by a Mellin transformation,
\beq
\label{gamP}
  \gamma_{\,\rm ik}^{\,(n)}(N,\als) \; = \;
  - \int_0^1 \!dx\:\, x^{\,N-1}\, P_{\,\rm ik}^{\,(n)}(x,\als) 
\:\: ,
\eeq
where the negative sign is a standard convention.
The splitting functions $P^{\,(n)}_{\rm ik}(x)$ are known to NNLO, i.e., 
at $n \leq 2$ in eq.~(\ref{Pexp}) \cite{MVV34}. 
The N$^3$LO contributions to eq.~(\ref{gamP}) have been obtained at 
$N \leq 6$ \cite{avRC17}; the results for $N = 2$ and $N = 4$ have been
presented in numerical form in ref.~\cite{avLL16}. 
Much more is known about their non-singlet counterparts \cite{DRUVV,MRUVV1},
see below.

Here we are interested in contributions with quartic color factors which 
we abbreviate~as
\beq
\label{d4abbr}
  d^{\,(4)}_{xy} \;\equiv\; d_x^{\,abcd} d_y^{\,abcd}
\; ,
\eeq
where $x,y$ labels the representations with generators $T_r^a$ and
\beq
\label{d4def}
  d_{r}^{\,abcd} \; =\; \frct{1}{6}\: {\rm Tr} \, ( \, 
   T_{r}^{a\,} T_{r}^{b\,} T_{r}^{c\,} T_{r}^{d\,}
   + \,\mbox{ five $bcd$ permutations}\, ) \; .
\eeq
In SU($n_c$), for fermions in the fundamental representation
(trace-normalized with $T_F = \frac{1}{2}$),
\bea
\label{d4AASUn}
  d^{\,(4)}_{A\!A}/n^{}_{\!A} &\!=\!& 
  \frct{1}{24}\: \ncs ( \ncs + 36 ) 
\; , \\
\label{d4FASUn}
  d^{\,(4)}_{F\!A}/n^{}_{\!A} &\!=\!& 
  \frct{1}{48}\: n_c ( \ncs + 6 ) 
\;, \\
\label{d4FFSUn}
  d^{\,(4)}_{F\!F}/n^{}_{\!A} &\!=\!&
  \frct{1}{96}\: ( \ncs - 6 + 18\:\! n_c^{\:\!-2} )
\; .
\eea
The dimension of the adjoint representation is related to 
$n^{}_{F} = n_c = C_A$ by $n^{}_{\!A} = (\ncs - 1) = 2\:\!n_c\:\! C_F $.
 
Terms with quartic Casimir invariants occur for the first time at four loops, 
the order considered here, in splitting functions, coefficient functions and
the beta function of QCD \cite{beta3}.
This effective `leading-order' situation implies particular relations and 
facilitates calculational simplifications.

The $d^{\,(4)}_{\rm xy}$ terms at four loops are scheme-independent. 
They should therefore fulfil the relation
\beq
\label{SUSYrel}
    \gamma_{\,\rm qq}^{\,(3)}(N) + \gamma_{\,\rm gq}^{\,(3)}(N)
  - \gamma_{\,\rm qg}^{\,(3)}(N) - \gamma_{\,\rm gg}^{\,(3)}(N)
  \;\:\Qeq\:\: 0
\eeq
for the color-factor substitutions \cite{avLL16}

\pagebreak
\vspace*{-9mm}

\beq
\label{SUSYcol}
      (2 \nf)^2 \, \frac{d^{\,(4)}_{F\!F}}{n_a}
 \,=\, 2 \nf \, \frac{d^{\,(4)}_{FA}}{n_a}
 \,=\, 2 \nf \, \frac{d^{\,(4)}_{F\!F}}{n_c}
 \,=\, \frac{d^{\,(4)}_{FA}}{n_c}
 \,=\, \frac{d^{\,(4)}_{AA}}{n_a}
\eeq
that lead to an ${\cal N}=1$ supersymmetric theory, for lower-order discussions
see refs.~\cite{SUSY23loop}. 
Here and below $\Qeq$ denotes equality for the quartic Casimir contributions.
The factor of two for each power of $\nf$ is due to the transition from QCD 
and its SU($n_c$) generalization to $\nf = 1$ Majorana fermions. 
We~have verified eq.~(\ref{SUSYrel}) at a sufficient number of $N$-values.
At higher values of $N$ this relation can be used to avoid the hardest 
diagram computations, those of the $d^{\,(4)}_{A\!A}$ contributions 
to~$\gamma_{\,\rm gg}^{\,(3)}$.

As in the non-singlet cases, the splitting functions are (conjectured to be)
constrained by a conformal symmetry of QCD at some non-integer space-time
dimension $D = 4 - 2 \ep$ \cite{BassoK06}. We find that the moments of the
off-diagonal splitting functions are consistent with the resulting prediction
in terms of reciprocity-respecting sums (see below), but fulfil the stronger, 
newly discovered condition
\beq
\label{Qoffd}
  \gamma_{\,\rm qg}^{\,(0)}(N) \,\gamma_{\,\rm gq}^{\,(3)}(N) 
  \;\:\Qeq\:\:
  \gamma_{\,\rm gq}^{\,(0)}(N) \,\gamma_{\,\rm qg}^{\,(3)}(N) 
\; .
\eeq
This result provides a stringent check of our very challenging high-$N$
computations. Other features resulting from the special status of 
quartic Casimir contributions at four loops are discussed below.

%
\setcounter{equation}{0}
\section{Diagram computations and $\bm N\!$-space results}
\label{sec:calcn}
\vspace*{-1mm}

Computations of four-loop inclusive DIS have been performed at $N\!\leq\!6$
for all colour factors in a manner analogous those at three loops in 
refs.~\cite{dis3moms}, for sample results see ref.~\cite{avLL16}. 
The ensuing moments of the four-loop splitting functions provide crucial
reference results for validating the present calculation performed in the 
framework of the operator-product expansion (OPE).
 
Our OPE diagram computations have been performed analogously to those 
presented in ref.~\cite{MRUVV1}.
The Feynman diagrams for the anomalous dimensions of the flavour-singlet
twist-2 spin-$N$ operators have been generated using QGRAF \cite{QGRAF}, and 
then processed by a {\sc Form} \cite{FORM} program, see ref.~\cite{jvLL2016}, 
that collects self-energy insertions, determines the colour factors and finds 
the topologies in the notation of the {\sc Forcer} package \cite{Forcer} that
performs the integral reduction after the harmonic projection \cite{MincForm}
to the desired value of $N$. For computational efficiency, diagrams with the 
same colour factor and topology are merged into meta-diagrams.

The main issue in these covariant-gauge calculations in a massless off-shell
case is the correct treatment of the gluon operators, 
see refs.~\cite{GluonOps}.
Since, as discussed above, we are dealing here with an effective lowest-order 
case, we are not confronted with the full complexity of this issue. 
We expect to return to this point in a future publication of all four-loop
contributions to the singlet anomalous dimensions. For three-loop on-shell 
OPE calculations with heavy quarks see ref.~\cite{BBK09}.

We now present our results for the quartic-Casimir contributions to the
anomalous dimensions (\ref{gamP}) at $N = 2, 4, \dots, 16$. These include
fractions and the values $\zeta_3$ and $\zeta_5$ of Riemann's $\zeta$-function,
but, as factorization-scheme independent `leading-order' contributions, do not 
include terms with even-$n$ values $\zeta_n^{}$, see refs.~\cite{no-pi2}. 
For brevity the results are written down in a numerical form.


The non-singlet and pure-singlet quark-quark anomalous dimensions include
\bea
\label{nsFA}
  \left. \gamma_{\,\rm ns}^{\,(3)+}(N) 
  \right|_{\,d^{\,(4)}_{F\!A}/n_F^{}} 
  &\!=\!& \mbox{} 
  + 773.10566 \: \dNnum2 \;\; 
  + 69.385963 \: \dNnum4 \;\;
  - 186.61376 \: \dNnum6 \;\;
\nn \\[-2mm] & &\mbox{}
  - 346.75182 \: \dNnum8 \;\;
  - 465.07282 \: \dNnum{10} \,
  - 559.57588 \: \dNnum{12}\;
\nn \\[0.4mm] & &\mbox{}
  - 638.52578 \: \dNnum{14}
  - 706.44946 \: \dNnum{16}\,
  -  \:\ldots \;\; , 
\\[2mm]
\label{nsFF}
  \left. \gamma_{\,\rm ns}^{\,(3)}(N) 
  \right|_{\,\nf\:\! d^{\,(4)}_{F\!F}/n_F^{}} \!\!
  &\!=\!& \mbox{} 
  - 65.736531 \: \dNnum2 \;\; 
  - 135.95246 \: \dNnum4 \;\;
  - 176.15626 \: \dNnum6 \;\;
\nn \\[-2mm] & &\mbox{}
  - 205.54604 \: \dNnum8 \;\;
  - 229.07719 \: \dNnum{10} \,
  - 248.80626 \: \dNnum{12} \;
\nn \\[0.4mm] & &\mbox{}
  - 265.82674 \: \dNnum{14} 
  - 280.80532 \: \dNnum{16}\, 
  -  \:\ldots \;\; , 
\\[2mm]
\label{psFF}
  \left. \gamma_{\,\rm ps}^{\,(3)}(N) 
  \right|_{\,\nf\:\! d^{\,(4)}_{F\!F}/n_F^{}} \!\!
  &\!=\!& \mbox{}
  - 146.97872\: \dNnum2 \;\; 
  - 88.852325\: \dNnum4 \;\;
  - 37.651992\: \dNnum6 \;\;
\nn \\[-2mm] & &\mbox{}
  - 20.290180\: \dNnum8 \;\;
  - 12.702121\: \dNnum{10} \,
  - 8.7568928\: \dNnum{12} \;
\nn \\[0.4mm] & &\mbox{}
  - 6.4465398\: \dNnum{14} 
  - 4.9735061\: \dNnum{16}\, 
  - \:\ldots
\;\; .
\eea
The corresponding results for the quark-gluon and gluon-quark quantities read
\bea
\label{gqFA}
  \left. \gamma_{\,\rm gq}^{\,(3)}(N)
  \right|_{\,d^{\,(4)}_{F\!A}/n_F^{}} \;\;
  &\!=\!& \mbox{} 
  - 773.10566 \: \dNnum2 \;\; 
  - 154.99156 \: \dNnum4 \;\;
  - 33.190677 \: \dNnum6 \;\;
\nn \\[-2mm] & &\mbox{}
  - 3.6877393 \: \dNnum8 \;\;
  + 5.6280884 \: \dNnum{10} \,
  + 8.8432407 \: \dNnum{12} \;
\nn \\[0.4mm] & &\mbox{}
  + 9.8467770 \: \dNnum{14} 
  +  \:\ldots \;\; ,
\\[2mm]
\label{gqFF}
  \left. \gamma_{\,\rm gq}^{\,(3)}(N)
  \right|_{\,\nf\:\! d^{\,(4)}_{F\!F}/n_F^{}} \!\!
  &\!=\!& \mbox{} 
  + 212.71525 \: \dNnum2 \;\; 
  + 40.812981 \: \dNnum4 \;\;
  + 20.540955 \: \dNnum6 \;\;
\nn \\[-2mm] & &\mbox{}
  + 13.623478 \: \dNnum8 \;\;
  + 10.207939 \: \dNnum{10} \,
  + 8.1771347 \: \dNnum{12} \;
\nn \\[0.4mm] & &\mbox{}
  + 6.8293658 \: \dNnum{14} 
  + 5.8681866 \: \dNnum{16}\, 
  +  \:\ldots 
\eea 
and 
\bea
\label{qgFA}
  \left. \gamma_{\,\rm qg}^{\,(3)}(N)
  \right|_{\,\nf\:\! d^{\,(4)}_{F\!A}/n_A^{}} \!\!
  &\!=\!& \mbox{} 
  - 386.55283 \: \dNnum2 \;\; 
  - 154.99156 \: \dNnum4 \;\;
  - 41.488346 \: \dNnum6 \;\;
\nn \\[-2mm] & &\mbox{}
  - 5.1628350 \: \dNnum8 \;\;
  + 8.4421327 \: \dNnum{10} \,
  + 13.896521 \: \dNnum{12} \;
\nn \\[0.4mm] & &\mbox{}
  + 16.001012 \: \dNnum{14} 
  +  \:\ldots \;\; ,
\\[2mm]
\label{qgFF}
  \left. \gamma_{\,\rm qg}^{\,(3)}(N)
  \right|_{\,\nfs\:\! d^{\,(4)}_{F\!F}/n_A^{}} \!\!
  &\!=\!& \mbox{} 
  + 106.35762 \: \dNnum2 \;\; 
  + 40.812981 \: \dNnum4 \;\;
  + 25.676194 \: \dNnum6 \;\;
\nn \\[-2mm] & &\mbox{}
  + 19.072869 \: \dNnum8 \;\;
  + 15.311908 \: \dNnum{10} \,
  + 12.849783 \: \dNnum{12} \;
\nn \\[0.4mm] & &\mbox{}
  + 11.097719 \: \dNnum{14} 
  + 9.7803110 \: \dNnum{16}\, 
  +  \:\ldots \;\; .
\eea
Finally the quartic-Casimir contributions to the four-loop gluon-gluon 
anomalous dimension are found to be 
\bea
\label{ggAA}
  \left. \gamma_{\,\rm gg}^{\,(3)}(N)
  \right|_{\,d^{\,(4)}_{A\!A}/n_A^{}} \;\;
  &\!=\!&
  \phantom{- 146.97872\: \dNnum2 \;\;}
  + 139.70415\: \dNnum4 \;\;
  - 42.404324\: \dNnum6 \;\;
\nn \\[-2mm] & &
  - 196.38527\: \dNnum8 \;\;
  - 317.16583\: \dNnum{10} \,
  - 414.93934\: \dNnum{12} \;
\nn \\[0.5mm] & &
  - 496.71624\: \dNnum{14}
  - \:\ldots \;\; ,
\\[2mm]
\label{ggFA}
  \left. \gamma_{\,\rm gg}^{\,(3)}(N)
  \right|_{\,\nf\:\! d^{\,(4)}_{F\!A}/n_A^{}} \!\!
  &\!=\!& \mbox{}
  + 386.55283 \: \dNnum2 \;\;
  - 441.47043 \: \dNnum4 \;\;
  - 458.54303 \: \dNnum6 \;\;
\nn \\[-2mm] & &\mbox{}
  - 462.79602 \: \dNnum8 \;\;
  - 469.79385 \: \dNnum{10} \,
  - 478.54929 \: \dNnum{12} \;
\nn \\[0.4mm] & &\mbox{}
  - 487.97876 \: \dNnum{14}
  - 497.49491 \: \dNnum{16}\,
  -  \:\ldots \;\; ,
\\[2mm]
\label{ggFF}
  \left. \gamma_{\,\rm gg}^{\,(3)}(N)
  \right|_{\,\nfs\:\! d^{\,(4)}_{F\!F}/n_A^{}} \!\!
  &\!=\!& \mbox{}
  - 106.35762 \: \dNnum2 \;\;
  - 117.11160 \: \dNnum4 \;\;
  - 121.74849 \: \dNnum6 \;\;
\nn \\[-2mm] & &\mbox{}
  - 123.79780 \: \dNnum8 \;\;
  - 124.86680 \: \dNnum{10} \,
  - 125.48946 \: \dNnum{12} \;
\nn \\[0.4mm] & &\mbox{}
  - 125.88108 \: \dNnum{14}
  - 126.14172 \: \dNnum{16}\,
  -  \:\ldots \;\; .
\eea
The absence of a $\dNnum2$ term, i.e., the vanishing of the $N\!=\!2$ 
contribution in eq.~(\ref{ggAA}) is required by the momentum sum rule. 
The implications of eqs.~(\ref{ggAA}) -- (\ref{ggFF}) for the large-$x$ limit
(\ref{xto1}) are addressed in section 4 below.

As for the non-singlet case in ref.~\cite{MRUVV1}, these fixed-$N$ results
are sufficient to deduce the all-$N$ form of the $\zeta_5$
contributions. The quark-quark anomalous dimensions can be expressed~as
\bea
\label{nsFAz5}
  \left. \gamma_{\,\rm ns}^{\,(3)+}(N)
  \right|_{\,\zr5\,d^{\,(4)}_{F\!A}/n_F^{}} 
  &\!=\!& 
  \frct{320}{3} \* \left(\,
    \S(1) \* ( 24\,\*\eta - 24\,\*\S(1) + 58 )
    - 69 \,\* \etaD2 + \frct{63}{2} \,\* \eta - 37 
  \right)
\; ,
\\[1mm]
\label{nsFFz5}
  \left. \gamma_{\,\rm ns}^{\,(3)}(N)
  \right|_{\,\zr5\,\nf\:\! d^{\,(4)}_{F\!F}/n_F^{}} \!\!
  &\!=\!& 
  \frct{1280}{3} \,\* \left(\,
    6 \,\* \etaD2 - 2 \,\* \S(1) - 5 \,\* \eta + 3
  \right)    
\; ,
\\[1mm]
\label{psFFz5}
  \left. \gamma_{\,\rm ps}^{\,(3)}(N)
  \right|_{\,\zr5\,\nf\:\! d^{\,(4)}_{F\!F}/n_F^{}} \!\!
  &\!=\!& 
  \frct{1280}{3} \* \left(\,
    9 \,\* \etaD2 + 14\,\* \eta - 4 \,\* \nu - \frct{1}{4} 
  \,\right)
\eea
in terms of the quantities
\bea
\label{etaDef}
  \eta &\!\equiv\!& \frct{1}{N} - \frct{1}{N+1} 
       \;\equiv\; \DNn{} - \DNp{}
       \;=\; \frct{1}{N(N+1)}
\:\: , \\
\label{nuDef}
  \nu &\!\equiv\!& \frct{1}{N-1} - \frct{1}{N+2}
      \;\equiv\; \DNm{} - \DNpp{}
      \;=\; \frct{3}{(N-1)(N+2)}  
\eea
and the harmonic sum $\S(1) \equiv \S(1)(N) = \sum_{\,k=1}^{\,N} k^{-1}$
which are reciprocity-respecting (RR), i.e., invariant under the replacement 
$N \ra 1-N$ corresponding to $f(x) \ra -x f(x^{\,-1})$ in $x$-space.

The all-$N$ results for the $\zr5\, d^{\,(4)}_{xy}$ contributions to 
the four-loop quark-gluon and gluon-quark anomalous dimensions read
\bea
\label{gqFAz5}
  \left. \gamma_{\,\rm gq}^{\,(3)}(N)
  \right|_{\,\zr5\,d^{\,(4)}_{F\!A}/n_F^{}} \;\;
  &\!=\!& 
  \frct{320}{3} \* \left(\,
    24 \,\* ( 2 \,\* \DNm{} - 2 \,\* \DNn{} + \DNp{} ) \,\* \S(1)
     - 24 \,\* \DNm2
     + 126 \,\* \DNn2
     + 63 \,\* \DNp2
\phantom{\frct{1}{2}} \right.  \nn \\[-2mm] & & \hspp \left. \mbox{}
     - 30 \,\* \DNm{}
     - 202 \,\* \DNn{}
     + \frct{391}{2} \,\* \DNp{}
     -  8 \,\* \DNpp{}
  \right)
\; , \\[1mm] 
\label{gqFFz5}
  \left. \gamma_{\,\rm gq}^{\,(3)}(N)
  \right|_{\,\zr5\,\nf\:\! d^{\,(4)}_{F\!F}/n_F^{}} \!\!
  &\!=\!&
 \frct{1280}{3} \* \left(\, 
     - 24 \,\* \DNn2
     - 12 \,\* \DNp2
     +  4 \,\* \DNm{}
     + 32 \,\* \DNn{}
     - 34 \,\* \DNp{}
 \right)
\eea 
and
\bea
\label{qgFAz5}
  \left. \gamma_{\,\rm qg}^{\,(3)}(N)
  \right|_{\,\zr5\,\nf\:\! d^{\,(4)}_{F\!A}/n_A^{}} \!\!
  &\!=\!& 
 \frct{640}{3} \* \left(\, 
    24 \,\* ( \DNn{} - 2 \,\* \DNp{} + 2 \,\* \DNpp{} ) \,\* \S(1)
     - 63 \,\* \DNn2
     - 126 \,\* \DNp2
     + 24 \,\* \DNpp2
\phantom{\frct{111}{2}} \right.  \nn \\[-2mm] & & \hspp \left. \mbox{}
     -  8 \,\* \DNm{}
     + \frct{391}{2} \,\* \DNn{}
     - 202 \,\* \DNp{}
     - 30 \,\* \DNpp{}
 \right)
\; , \\[1mm]
\label{qgFFz5}
  \left. \gamma_{\,\rm qg}^{\,(3)}(N)
  \right|_{\,\zr5\,\nfs\:\! d^{\,(4)}_{F\!F}/n_A^{}} \!\!
  &\!=\!& \mbox{}
 \frct{2560}{3} \* \left(\, 
       12 \,\* \DNn2
     + 24 \,\* \DNp2
     - 34 \,\* \DNn{}
     + 32 \,\* \DNp{}
     +  4 \,\* \DNpp{}
 \right)
\; .
\eea
By multiplying eqs.~(\ref{gqFAz5}) and (\ref{gqFFz5}) with 
\beq
  \gamma_{\,\rm qg}^{\,(0)}(N) \:=\: 
  - 2\:\!C_F\, (\DNn{}- 2\:\!\DNp{}+ 2\:\!\DNpp{}) \; ,
\eeq
and eqs.~(\ref{qgFAz5}) and (\ref{qgFFz5}) with
\beq
  \gamma_{\,\rm gq}^{\,(0)}(N) \:=\: 
  - 2\:\!\nf\, (2\:\!\DNm{}- 2\:\!\DNn{}+ \DNp{}) \; ,
\eeq
one arrives at two RR expressions that fulfil eq.~(\ref{Qoffd}) at all $N$; 
the required relation between $n_A^{}$, $n_F^{}$ and $C_F$ has been given 
below eq.~(\ref{d4FFSUn}).

The corresponding gluon-gluon anomalous dimension are given by the RR 
expressions
\bea
\label{ggAAz5}
  \left. \gamma_{\,\rm gg}^{\,(3)}(N)
  \right|_{\,\zr5\,d^{\,(4)}_{A\!A}/n_A^{}} \;\;
  &\!=\!&
 \frct{64}{3} \* \left(\, 
    30 \* \left( 12 \,\* \etaD2 - 4 \,\* \nuD2 
       - \,\* \S(1) \* ( 4\,\*\S(1) + 8 \,\* \eta - 8\,\* \nu - 11 ) 
       - 7 \* \nu \right) 
\phantom{\frct{1}{2}} \right.  \nn \\[-2mm] & & \hspp \left. \mbox{}
       + 188 \,\* \eta - \frct{751}{3} - \frct{1}{6}\,\* N\,\*(N+1)
 \right)
\; , \\[1mm]
\label{ggFAz5}
  \left. \gamma_{\,\rm gg}^{\,(3)}(N)
  \right|_{\,\zr5\,\nf\:\! d^{\,(4)}_{F\!A}/n_A^{}} \!\!
  &\!=\!& 
 \frct{128}{3} \* \left(\, 
    10 \* \left( 15 \,\* \etaD2 - 6 \,\*\S(1) + 2 \,\*\nu \right)
    - 121 \,\* \eta + \frct{287}{3} + \frct{1}{3}\,\* N\,\*(N+1)
 \right)
\; , \quad \\[2mm]
\label{ggFFz5}
  \left. \gamma_{\,\rm gg}^{\,(3)}(N)
  \right|_{\,\zr5\,\nfs\:\! d^{\,(4)}_{F\!F}/n_A^{}} \!\!
  &\!=\!& 
 \frct{256}{3} \* \left(\, 
    - 120 \,\* \etaD2 + 23 \,\* \eta - \frct{17}{6} 
    - \frct{1}{6}\,\* N\,\*(N+1)
 \right)
\; .
\eea
These results exhibit interesting features in the large-$N$ threshold limit
and the $N \!\ra\! 1$ BFKL limit.


Unlike all $N$-space expressions for QCD splitting functions calculated up
to now, eqs.~(\ref{ggAAz5}) -- (\ref{ggFFz5}) include terms of the form
$\,\zr5\, N (N+1)\,$. In the complete results, these terms have to be 
compensated by contributions that develop $\zr5$-terms in the limit  
$N \!\ra\! \infty$, since the overall leading large-$N$ behaviour is given by
$\ln N$ multiplied by the cusp anomalous dimension \cite{Korch89} due to 
the Mellin transform of eq.~(\ref{xto1}). 
This compensation has occurred before, in the three-loop coefficient functions 
for inclusive DIS \cite{MVV610+}, where $\zr5$ enters with positive powers 
of $N$ in the combination
\beq
\label{fNfct}
     f(N) \;=\;
        5 \* \zr5
       - 2 \* \S(-5)
       + 4 \:\!\* \zr3 \:\!\* \S(-2) 
       - 4 \:\!\* \Ss(-2,-3)
       + 8 \* \Sss(-2,-2,1)
       + 4 \:\!\* \Ss(3,-2)
       - 4 \:\!\* \Ss(4,1)
       + 2 \* \S(5)
\eeq
of $\zeta$-values and harmonic sums \cite{HSums} that ensures the
correct large-$N$ behaviour. It may be worthwhile to note that the
$N (N+1)$ terms in eqs.~(\ref{ggAAz5}) -- (\ref{ggFFz5}) cancel in the
SUSY limit (\ref{SUSYcol}).

In addition, both eq.~(\ref{nsFAz5}) and (\ref{ggAAz5}) include terms of
the form $\zr5 [S_1(N)]^2$
-- with the same coefficients, as required in view of eqs.~(\ref{SUSYrel})
and (\ref{SUSYcol}) -- that also need to be compensated in the large-$N$
limit. A natural possibility is that the diagonal QCD splitting
function include terms with $[S_1(N)]^2\, f(N)$, i.e., the same structure as 
the `wrapping correction' in the anomalous dimensions in $\,{\cal N}=4$ 
maximally supersymmetric Yang-Mills theory \cite{N=4wrap}.
Unfortunately we are not (yet) in a position to derive the all-$N$ structure
of the $\zr3$-terms, which could provide further evidence for (or~exclude) 
the occurrence of the function (\ref{fNfct}) in the four-loop anomalous 
dimensions in~QCD.

In the limit $N\!\ra\! 1$, eqs.~(\ref{gqFAz5}) and (\ref{ggAAz5}) include terms
with $1/(N-1)^2$. Since the leading terms at four-loop are proportional to
$1/(N-1)^4$ \cite{BFKL2}, these represent next-to-next-to-leading logarithmic
(NNLL) contributions in this high-energy (small-$x$) limit.
Unless these terms are compensated by contributions that develop $\zr5$-terms
in the limit $N \!\ra\! 1$, the complete NNLL four-loop contributions in QCD 
cannot possibly be obtained by resumming lower-order information, as such 
information cannot predict coefficients of quartic Casimir invariants.

%
\setcounter{equation}{0}
\section{$\bm x$-space results and cusp anomalous dimensions}
\label{sec:cusp}
\vspace*{-1mm}

The fixed-$N$ moments (\ref{nsFA}) -- (\ref{ggFA}) of the quartic-Casimir
contributions to the four-loop splitting functions can be employed to obtain 
$x$-space approximations which small uncertainties at least at $x \gsim 0.1$. 
In~the quark-quark and gluon-gluon cases, 
these approximations involve only two unknown coefficients of terms that do 
not vanish for $x \ra 1$, i.e., the coefficients $A_4$ and $B_4$ in 
eq.~(\ref{xto1}). The predictable coefficients $C_4$ and $D_4$ vanish for the 
$d^{\,(4)}_{xy}$ terms, since there are no lower-order quantities with these
colour factors. Consequently, the coefficients of the leading large-$x$ terms, 
i.e., the cusp anomalous dimensions, can be determined with a rather high 
accuracy.

This programme has been carried out in ref.~\cite{MRUVV1} for the complete
non-leading large-$n_c$ (N$n_c$) $\nfz$ and $\nfo$ parts of 
$P_{\rm ns}^{\,(3)}(x)$ in QCD as well as for all individual colour factors. 
The leading large-$n_c$ contributions and the $\nfs$ and $\nft$ terms are
completely known \cite{DRUVV,MRUVV1,JAG94}.
The results for $A_{4,\rm q}$ are collected in table~1, where the $\nfz$ part 
has been improved upon using the N$n_c$ results for QCD. 
The~coefficients of $A_{4,\rm q}$ which are known exactly have also been 
determined from the quark form factor \cite{FFnf2,qFF}, the results are in 
complete agreement.  Very recently, the exact coefficient of $\cft \nf$ has 
been obtained in ref.~\cite{Grozin18}.

\begin{table}[b!]
\vspace*{-1mm}
\centering
  \renewcommand{\arraystretch}{1.2}
  \begin{tabular}{MMMM}
   \mbox{quark}  & \mbox{gluon} &    A_{4,\rm q}             &  A_{4,\rm g}     \\[1pt]
   \hline\\[-3.5mm]
    \cff         &     -        &     0                      &        -              \\[1pt]
    \cft\, \ca   &     -        &     0                      &        -              \\[1pt]
    \cfs \cas    &     -        &     0                      &        -              \\[1pt]
    \cf \cat     &   \caf       & \phantom{-}  610.25 \pm 0.1~~ &                    \\[1pt]
    d^{\,(4)}_{F\!A}/N_{\!F}^{} & d^{\,(4)}_{A\!A}/N_{\!A}^{} 
                                &    -507.0 \pm 2.0          & -507.0 \pm 5.0~~      \\[0.5mm]
\hline\\[-3.5mm]
    \nf\, \cft   & \nf\,\cfs\ca &    -31.00554               &                          \\[1pt]
  \nf\,\cfs\ca   & \nf\,\cf\cas & \phantom{-} 38.75 \pm 0.2  &                          \\[1pt]
  \nf\,\cf\cas   & \nf\cat      &  -440.65 \pm 0.2~~         &                          \\[1pt]
    \nf\,d^{\,(4)}_{F\!F}/N_{\!F}^{} & \nf\,d^{\,(4)}_{F\!A}/N_{\!A}^{}   
                                &  -123.90  \pm 0.2~~        & -124.0 \pm 0.6~~         \\[0.5mm]
\hline\\[-3.5mm]
    \nfs\,\cfs   & \nfs\,\cf\ca &   -21.31439                &                          \\[-0.3mm]
  \nfs\,\cf\ca   & \nfs\,\cas   & \phantom{-}58.36737        &                          \\[-0.3mm]
       -         & \nfs\, d^{\,(4)}_{F\!F}/N_{\!A}^{}  
                                &      -                     & ~~\phantom{-}0.0 \pm 0.1 \\[-0.3mm]
    \nft\,\cf    & \nft\,\ca    & \phantom{-}2.454258        & ~~\phantom{-}2.454258    \\[0.2mm]
  \hline
  \end{tabular}
  \vspace*{1mm}
  \caption{\small \label{tab:AB}
  Fourth-order coefficients of the quark and gluon cusp anomalous dimensions
  determined from the large-$x$ limit (\ref{xto1}) of the quark-quark and
  gluon-gluon splitting functions. The errors in the quark case are correlated
  due to the exactly known large-$n_c$ limit.
  Our numerical value of $-31.00 \pm 0.4$ \cite{MRUVV1} for the coefficient of 
  $\nf \cft$ in $A_{4,\rm q}$ has been replaced by recent exact result of 
  ref.~\cite{Grozin18}. 
  This and the exact values for the $\nfs$ and $\nft$ coefficients have been 
  rounded to seven digits.
  }
  \vspace*{-2mm}
\end{table}

We have now performed analogous determinations of the quartic-Casimir
coefficients of $A_{4,\rm g}$. The results are also shown in table~1, together
with the only piece known exactly so far, the $C_A \nft$ contribution
\cite{DRUVV,gFFnf3}. We see that, as up to the third order \cite{MVV34},
the corresponding quark and gluon entries have the same coefficients
(for now: as far as they have been computed, and within numerical errors).
We refer to this (for now: conjectured) relation as {\em generalized Casimir
scaling}.

Unlike to three loops, this relation does not have the consequence that the
values of $A_{4,\rm g}$ and $A_{4,\rm q}$ are related by a simple numerical 
Casimir scaling in QCD, i.e., a factor of $C_A / C_F = 9/4$. 
However, this numerical Casimir scaling is restored in the large-$n_c$ limit 
of the quartic colour factors, and therefore also in the overall large-$n_c$ 
limit, see also ref.~\cite{Dixon17}.

The results of refs.~\cite{DRUVV,MRUVV1} and the present paper lead to the
following results for the four-loop quark and gluon cusp anomalous dimensions,
expanded in powers of $\als/(4\pi)$, recall eq.~(\ref{Pexp}),
\bea
\label{A4q}
  A_{4,\rm q} &\,=\,&   20702(2)\phantom{0} \,-\, 5171.9(2) \,\nf
                  \,+\, 195.5772\,\nfs      \,+\, 3.272344 \,\nft
\; , \\[-0.5mm]
\label{A4g}
  A_{4,\rm g} &\,=\,&   40880(30)      \,-\, 11714(2) \;\nf\,
                  \,+\, 440.0488\,\nfs \,+\, 7.362774 \,\nft
\; .
\eea
For comparison, the large-$n_c$ coefficients of $A_{4,\rm q}$ (not changing
the overall factor of $\cf$) read 21209.0, 5179.37 and 190.841, respectively,
for the $\nfz$, $\nfo$ and $\nfs$ contributions. The numerical Casimir scaling
between $A_{4,\rm g}$ and $A_{4,\rm q}$ is broken by almost 15\% in the $\nfz$
terms.
This breaking is due to the non-leading large-$n_c$ (N$n_c$) part of the 
quartic-Casimir term, which is larger by a factor of 6 in $A_{4,\rm g}$ 
than in $A_{4,\rm q}$ due to `36' and `6' in eqs.~(\ref{d4AASUn}) and 
(\ref{d4FASUn}). 
This much larger size of the N$n_c$ contribution in the gluon case also leads
to the much larger uncertainty of its $\nfz$ coefficient.

Combining the above with the lower-order coefficients 
, we arrive at the very benign expansions
\bea
\label{Aqnf345}
  A_{\rm q}(\als,\nf\!=\!3) &\!=\!& 0.42441\,\als \,
  [\, 1 +  0.72657\, \als +  0.73405\, \as(2) + 0.6647(2)\, \as(3) + \ldots\, ]
\; , \nn \\[-0.2mm]
  A_{\rm q}(\als,\nf\!=\!4) &\!=\!& 0.42441\,\als \,
  [\, 1 +  0.63815\, \als +  0.50998\, \as(2) + 0.3168(2)\, \as(3) + \ldots\, ]
\; , \nn \\[-0.2mm]
  A_{\rm q}(\als,\nf\!=\!5) &\!=\!& 0.42441\,\als \,
  [\, 1 +  0.54973\, \als +  0.28403\, \as(2) + 0.0133(3)\, \as(3) + \ldots\, ]
\eea
and
\bea
\label{Agnf345}
  A_{\rm g}(\als,\nf\!=\!3) &\!=\!& 0.95493\,\als \,
  [\, 1 + 0.72657\, \als  +  0.73405\, \as(2) + 0.415(2)\, \as(3) + \ldots\, ]
\; , \nn \\[-0.2mm]
  A_{\rm g}(\als,\nf\!=\!4) &\!=\!& 0.95493\,\als \,
  [\, 1 + 0.63815\, \als  +  0.50998\, \as(2) + 0.064(2)\, \as(3) + \ldots\, ]
\; , \nn \\[-0.2mm]
  A_{\rm g}(\als,\nf\!=\!5) &\!=\!& 0.95493\,\als \,
  [\, 1 + 0.54973\, \als  +  0.28403\, \as(2) - 0.243(2)\, \as(3) + \ldots\, ]
\eea
in terms of $\als$
for the physically relevant values of the number $\nf$ of light flavours.
Due to the additional cancellations between the terms without and with $\nf$
in eq.~(\ref{A4q}) and (\ref{A4g}), the numerical Casimir scaling is 
completely broken in fourth-order contributions.
 
The remaining uncertainties are practically irrelevant for all phenomenological
applications, which include (but are by no means exhausted by) calculations of 
the soft-gluon exponentiation at  next-to-next-to-next-to-leading logarithmic 
(N$^3$LL) \cite{n3llSGE} and higher accuracy, see, e.g., ref.~\cite{dFMMV}, 
and similar calculations in other frameworks such as soft-collinear 
effective theory.
 
Another application is the absolute ratio
$|{\cal F}^{\! g}(q^2)/{\cal F}^{\! g}(-q^2)|$ of the renormalized time-like
and space-like Higgs-gluon-gluon form factors in the heavy-top limit. 
This quantity is infrared finite and directly enters the cross section for 
Higgs boson production in hadronic collisions. 
Using eq.~(\ref{Agnf345}) for the small $A_{4,\rm g}$ contribution, we can 
update the result of ref.~\cite{MVV9} which used a value based on a Pad\'e 
estimate for $A_{4,\rm q}$ and numerical Casimir scaling. 
The new result for $n_{\!f} = 5$ reads
\beq
\label{HggTS}
  \bigg| \frct{{\cal F}^{\! g}(q^2)}{{\cal F}^{\! g}(-q^2)} \bigg|^2 
  \: = \:
        1 + 4.7124\: \als   + 13.694\: \as(2)
          + 25.935\: \as(3) + (34.82 \pm 0.01)\: \as(4) 
          \:+\: \ldots
\; .
\eeq
While the coefficient of $\as(4)$ is noticeably smaller than in 
ref.~\cite{MVV9}, the general pattern is unchanged: large coefficients, 
but definitely no sign of a runaway growth -- on the contrary.
The numerical $\as(4)$ effect in eq.~(\ref{HggTS}) is a fraction of a 
percent at scales close to the mass of the Higgs boson.

%
\setcounter{equation}{0}
\section{Summary}
\label{sec:summ}
\vspace*{-1mm}

We have presented the first calculations of a substantial number of moments 
of contributions to the four-loop (N$^3$LO) flavour-singlet splitting functions 
$P_{\rm ik}^{\,(3)}$ outside the large-$\nf$ limit. 
Specifically, we have obtained the even moments $N\!\leq\!16$ of all terms 
with quartic Casimir invariants.
The~calculations have been performed in the framework of the operator-product
expansion; the results at $N\leq 6$ (and partly at $N=8$) have been 
checked against those of conceptually much simpler, but computationally much 
harder determinations via structure functions in deep-inelastic scattering.

Our results show features expected for these effectively lowest-order 
contributions, such as the supersymmetric relation, and properties not 
predicted before, in particular a simple relation between the quartic-Casimir
parts of the off-diagonal splitting functions $P_{\rm qg}^{\,(3)}$ and 
$P_{\rm gq}^{\,(3)}$.
We have obtained the all-$N$ expressions for the $\zeta_5$ parts.
The diagonal quantities $P_{\rm qq}^{\,(3)}$ and $P_{\rm gg}^{\,(3)}$ include
contributions which have the structure of the wrapping corrections found in 
${\cal N} \!=\! 4$ maximally supersymmetric Yang-Mills theory.
The all-$N$ expressions for $P_{\rm gg}^{\,(3)}$ includes numerator-$N$ terms.
Such terms are not entirely new, but have not been encountered in splitting 
functions before.

The calculated moments of $P_{\rm gg}^{\,(3)}$ enable a numerical
determination of a quantity that is important in a much wider context, the 
(light-like) four-loop gluon cusp anomalous dimension $A_{4,\rm g}$.
We~find for the quartic-Casimir parts, and conjecture for all other terms,
that the coefficients for $A_{4,\rm g}$ are related to those of its quark
counterpart $A_{4,\rm q}$ by a direct generalization of the Casimir scaling 
found at lower orders. 
This allows us to present numerical results for $A_{4,\rm g}$ in QCD that 
are sufficiently accurate for all phenomenological purposes.
Due to differences in (the contributions that are non-leading in the
large-$n_c$ limit to) the quartic colour factors, there is no simple
relation between the numerical values of $A_{4,\rm g}$ and $A_{4,\rm q}$ 
for physical values of the number of flavours $\nf$.

{\sc Form} files of our fixed-$N$ and all-$N$ moments of the four-loop
splitting functions, including the analytic expressions for the former 
quantities not shown in section 3, can be obtained from the preprint server 
https://arXiv.org by downloading the source of this article.
They are also available from the authors upon request.

%
\subsection*{Acknowledgements}
\vspace*{-1mm}
A.V. would like to thanks E. Gardi for useful discussions on the diagrammatic
basis of the generalized Casimir scaling.
This work has been supported by the {\it Deutsche Forschungsgemeinschaft} 
(DFG) under grant number MO~1801/2-1, and by the Advanced Grant 320651,
{\it HEPGAME}, of the {\it European Research Council}$\,$ (ERC). 

{\footnotesize


\begin{thebibliography}{100}

\bibitem{MVV610+}
J.A.M.~Vermaseren, A.~Vogt and S.~Moch,
  Nucl.\ Phys.\ B724 (2005) 3, hep-ph/0504242; \\[0.2mm]
S. Moch, J.A.M. Vermaseren, A.~Vogt,
  Nucl. Phys. B813 (2009) 220, arXiv:0812.4168; \\[0.2mm]
J.~Davies, A.~Vogt, S.~Moch and J.A.M.~Vermaseren,
  PoS (DIS$\,$2016) 059, arXiv:1606.08907

\bibitem{Higgs1}
C.~Anastasiou, C.~Duhr, F.~Dulat, F.~Herzog and B.~Mistlberger,
  Phys.\ Rev.\ Lett.\ 114 (2015) 212001, arXiv:1503.06056;~ 
C.~Anastasiou et al.,
  JHEP 05 (2016) 058, arXiv:1602.00695; \\[0.2mm]
B.~Mistlberger,
  arXiv:1802.00833

\bibitem{Higgs2}
F.A.~Dreyer and A.~Karlberg,
  Phys.\ Rev.\ Lett.\ 117 (2016) 072001, arXiv:1606.00840

\bibitem{N3LOjet}
J.~Currie, T.~Gehrmann, E.W.N.~Glover, A.~Huss, J.~Niehues, A.~Vogt,
  arXiv:1803.09973 (JHEP, to appear)

\bibitem{Korch89}
G.~P.~Korchemsky,
  Mod.\ Phys.\ Lett.\ A4 (1989) 1257

\bibitem{DRUVV}
J.~Davies, B.~Ruijl, T.~Ueda, J.A.M. Vermaseren, A. Vogt,
  Nucl.\ Phys.\ B915 (2017) 335,~arXiv:1610.07477

\bibitem{MRUVV1}
S.~Moch, B.~Ruijl, T.~Ueda, J.A.M.~Vermaseren and A.~Vogt,
  JHEP 10 (2017) 041, arXiv:1707.08315

\bibitem{OffDxto1}
A.A.~Almasy, G.~Soar and A.~Vogt,
  JHEP 03 (2011) 030, arXiv:1012.3352; \\[0.2mm]
A.A.~Almasy, N.A.~Lo Presti and A.~Vogt,
  JHEP 01 (2016) 028, arXiv:1511.08612

\bibitem{DMS05}
Y.L.~Dokshitzer, G.~Marchesini and G.P.~Salam,
  Phys.\ Lett.\ B634 (2006) 504, hep-ph/0511302

\bibitem{BFKL1}
E.A. Kuraev, L.N. Lipatov and V.S. Fadin,
  Sov.\ Phys.\ JETP 45 (1977) 199; \\[0.2mm]
I.I. Balitsky and L.N. Lipatov,
  Sov.\ J. Nucl.\ Phys.\ 28 (1978) 822

\bibitem{BFKL2}
T. Jaroszewicz,
  Phys.\ Lett.\ B116 (1982) 291; \\[0.2mm]
S. Catani, F. Fiorani and G. Marchesini,
  Nucl.\ Phys.\ B336 (1990) 18; \\[0.2mm]
S.~Catani and F.~Hautmann,
  Nucl.\ Phys.\ B427 (1994) 475, hep-ph/9405388

\bibitem{MVV34}
S. Moch, J.A.M. Vermaseren and A. Vogt,
  Nucl. Phys. B688 (2004) 101, hep-ph/0403192; \\[0.2mm]
A. Vogt, S. Moch and J.A.M. Vermaseren,
  Nucl. Phys. B691 (2004) 129, hep-ph/0404111

\bibitem{avRC17}
A.~Vogt, S.~Moch, B.~Ruijl, T.~Ueda and J.~Vermaseren,
  PoS ({\sc Radcor}$\,$2017) 046, arXiv:1801.06085 

\bibitem{avLL16}
B.~Ruijl, T.~Ueda, J.A.M.~Vermaseren, J.~Davies and A.~Vogt,
  PoS (LL$\,$2016) 071, arXiv:1605.08408

\bibitem{beta3}
T.~van Ritbergen, J.A.M.~Vermaseren and S.A.~Larin,
  Phys.\ Lett.\ B400 (1997) 379, hep-ph/9701390; \\[0.2mm]
M.~Czakon, 
  Nucl.\ Phys.\ B710 (2005) 485, hep-ph/0411261

\bibitem{SUSY23loop}
I.~Antoniadis and E.~G.~Floratos,
  Nucl.\ Phys.\ B191 (1981) 217; \\[0.2mm]
A.A.~Almasy, S.~Moch and A.~Vogt,
  Nucl.\ Phys.\ B854 (2012) 133, arXiv:1107.2263

\bibitem{BassoK06}
B.~Basso and G.~P.~Korchemsky,
  Nucl.\ Phys.\ B775 (2007) 1, hep-th/0612247

\bibitem{dis3moms}
S.A.~Larin, T.~van Ritbergen and J.A.M.~Vermaseren,
  Nucl.\ Phys.\ B427 (1994) 41; \\[0.2mm]
S. Larin, P. Nogueira, T. van Ritbergen, J. Vermaseren,
  Nucl.\ Phys.\ B492 (1997) 338, hep-ph/9605317

\bibitem{QGRAF}
P.~Nogueira,
  J. Comput.\ Phys.\ 105 (1993) 279

\bibitem{FORM}
J.A.M.~Vermaseren, math-ph/0010025; \\[0.2mm]
M.~Tentyukov and J.A.M.~Vermaseren,
  Comput.\ Phys.\ Commun.\  181 (2010) 1419, hep-ph/0702279; \\[0.2mm]
J.~Kuipers, T.~Ueda, J.A.M.~Vermaseren and J.~Vollinga,
  CPC 184 (2013) 1453, arXiv:1203.6543; \\[0.2mm]
B.~Ruijl, T.~Ueda and J.A.M.~Vermaseren, arXiv:1707.06453

\bibitem{jvLL2016}
  F.~Herzog, B.~Ruijl, T.~Ueda, J.A.M.~Vermaseren and A.~Vogt,
  PoS (LL$\,$2016) 073, arXiv:1608.01834

\bibitem{Forcer}
B.~Ruijl, T.~Ueda and J.A.M.~Vermaseren,
  arXiv:1704.06650 

\bibitem{MincForm}
S.A.~Larin, F.V.~Tkachov and J.A.M.~Vermaseren,
  NIKHEF-H-91-18

\bibitem{GluonOps}
R.~Hamberg and W.~L.~van Neerven,
  Nucl.\ Phys.\ B379 (1992) 143; \\[0.2mm]
J.C.~Collins and R.J.~Scalise,
  Phys.\ Rev.\ D50 (1994) 4117, hep-ph/9403231; \\[0.2mm]
B.W.~Harris and J.~Smith,
  Phys.\ Rev.\ D51 (1995) 4550, hep-ph/9409405; \\[0.2mm]
Y.~Matiounine, J.~Smith and W.L.~van Neerven,
  Phys.\ Rev.\ D57 (1998) 6701, hep-ph/9801224

\bibitem{BBK09}
I.~Bierenbaum, J.~Bl{\"u}mlein and S.~Klein, 
  Nucl.\ Phys.\ B820 (2009) 417, arXiv:0904.3563

\bibitem{no-pi2}
M.~Jamin and R.~Miravitllas,
  Phys.\ Lett.\ B779 (2018) 452, arXiv:1711.00787; \\[0.2mm]
J. Davies and A. Vogt,
  Phys.\ Lett.\ B776 (2018) 189, arXiv:1711.05267; \\[0.2mm]
P.A.~Baikov and K.G.~Chetyrkin,
  arXiv:1804.10088 

\bibitem{HSums}
J.A.M. Vermaseren,
  Int.\ J.\ Mod.\ Phys.\ A14 (1999) 2037, hep-ph/9806280;\\[0.2mm]
J. Bl\"umlein and S. Kurth,
  Phys.\ Rev.\ D60 (1999) 014018, hep-ph/9810241

\bibitem{N=4wrap}
A.V.~Kotikov, L.N.~Lipatov, A.~Rej, M.~Staudacher and V.N.~Velizhanin,
  J.\ Stat.\ Mech.\ 10 (2007) P10003, arXiv:0704.3586;~
Z.~Bajnok, R.A.~Janik and T.~Lukowski, 
  Nucl.~Phys.\ B816 (2009) 376, arXiv:0811.4448

\bibitem{JAG94}
J.A.~Gracey,
  Phys.\ Lett.\ B322 (1994) 141, hep-ph/9401214

\bibitem{FFnf2}
A.~Grozin, J.M.~Henn, G.P.~Korchemsky and P.~Marquard,
  JHEP 1601 (2016) 140, arXiv:1510.07803;\\[0.2mm]
A.~Grozin,  PoS (LL$\,$2016) 053, arXiv:1605.03886;\\[0.2mm]
R.N.~Lee, A.V.~Smirnov, V.A.~Smirnov and M.~Steinhauser,
  Phys.\ Rev.\ D96 (2017) 014008, arXiv:1705.06862

\bibitem{qFF}
J.~Henn, A.V.~Smirnov, V.A.~Smirnov and M.~Steinhauser,
  JHEP 1605 (2016) 066, arXiv:1604.03126; \\[0.2mm]
J.~Henn, A.V.~Smirnov, V.A.~Smirnov, M.~Steinhauser and R.N.~Lee,
  JHEP 03 (2017) 139, arXiv:1612.04389

\bibitem{Grozin18}
A.~Grozin,
  arXiv:1805.05050

\bibitem{gFFnf3}
A.~von Manteuffel and R.~M.~Schabinger,
  Phys.\ Rev.\ D95 (2017) 034030, arXiv:1611.00795

\bibitem{Dixon17}
L.J.~Dixon,
  JHEP 01 (2018) 075, arXiv:1712.07274

\bibitem{n3llSGE}
S.~Moch, J.A.M.~Vermaseren and A.~Vogt,
  Nucl.\ Phys.\ B726 (2005) 317, hep-ph/0506288; \\[0.2mm]
S.~Moch and A.~Vogt,
  Phys.\ Lett.\ B631 (2005) 48, hep-ph/0508265; \\[0.2mm]
S.~Moch and A.~Vogt,
  Phys.\ Lett.\ B680 (2009) 239, arXiv:0908.2746

\bibitem{dFMMV}
D.~de Florian, J.~Mazzitelli, S.~Moch and A.~Vogt,
  JHEP 10 (2014) 176, arXiv:1408.6277

\bibitem{MVV9}
S.~Moch, J.A.M.~Vermaseren and A.~Vogt,
  Phys.\ Lett.\ B625 (2005) 245, hep-ph/0508055

\end{thebibliography}

\providecommand{\href}[2]{#2}\begingroup\raggedright\endgroup

}

\end{document}